\documentclass[eps,twocolumn,preprintnumbers]{revtex4}
\usepackage{graphics}
\usepackage{graphicx}
\usepackage{dcolumn} 
\usepackage{bm}
\usepackage{color}
\usepackage{epsfig}
\usepackage{comment}
\newcommand{\cca}{Cr$_2$CoAl}

\begin{document}
\title{Noncentrosymmetric compensated half-metal hosting\\ 
  pure spin Weyl nodes, triple nodal points, nodal loops, and nexus fermions} 
\author{Hyo-Sun Jin$^1$}
\author{Young-Joon Song$^1$}
\author{Warren E. Pickett$^2$}
\email{pickett@physics.ucdavis.edu}
\author{Kwan-Woo Lee$^{1,3,4}$}
\email{mckwan@korea.ac.kr}
\affiliation{
 $^1$Department of Applied Physics, Graduate School, Korea University, Sejong 30019, Korea\\
 $^2$Department of Physics, University of California, Davis, California 95616, USA\\
 $^3$IFW Dresden, Helmholtzstr. 20, D-01069, Dresden, Germany\\
 $^4$Division of Display and Semiconductor Physics, Korea University, Sejong 30019, Korea
}
\date{\today}
\begin{abstract}
Materials containing multiple topological characteristics become more exotic 
when combined with noncentrosymmetric crystal structures and unusual magnetic phases
such as the compensated half-metal state, which is gapped in one spin direction and
conducting in the other. First principles calculations 
reveal these multiple topological features in the compensated half-metal \cca~ 
having neither time-reversal nor inversion symmetries.
In the absence of (minor) spin-orbit coupling (SOC), there are
(1) a total of twelve pairs of magnetic Weyl points, 
(2)  three distinct sets of triple nodal points (TNPs)
    near the Fermi level that are 
(3) interconnected with six symmetry related nodal lines. 
This combination gives rise to
{\it fully spin polarized nexus fermions}, in a system with broken time-reversal
symmetry but negligible macroscopic magnetic field.
The observed high Curie temperature of 750 K and calculated SOC hybridization
mixing of several meV should make these nexus fermions readily measurable.
Unlike topological features discussed for other Heuslers which emphasize their
strong ferromagnetism, this compensated half-metal is impervious to typical magnetic
fields, thus providing a complementary set of experimental phenomena. 
Making use of the soft calculated magnetic state, large magnetic fields can be used
to rotate the direction of magnetism, during which certain topological 
features will evolve.
Our results suggest that these features may be common
in inverse-Heusler systems, particularly the isostructural and isovalent Ga and In analogs.
\end{abstract}
\maketitle


\begin{figure*}[tbp]
{\resizebox{16cm}{4.2cm}{\includegraphics{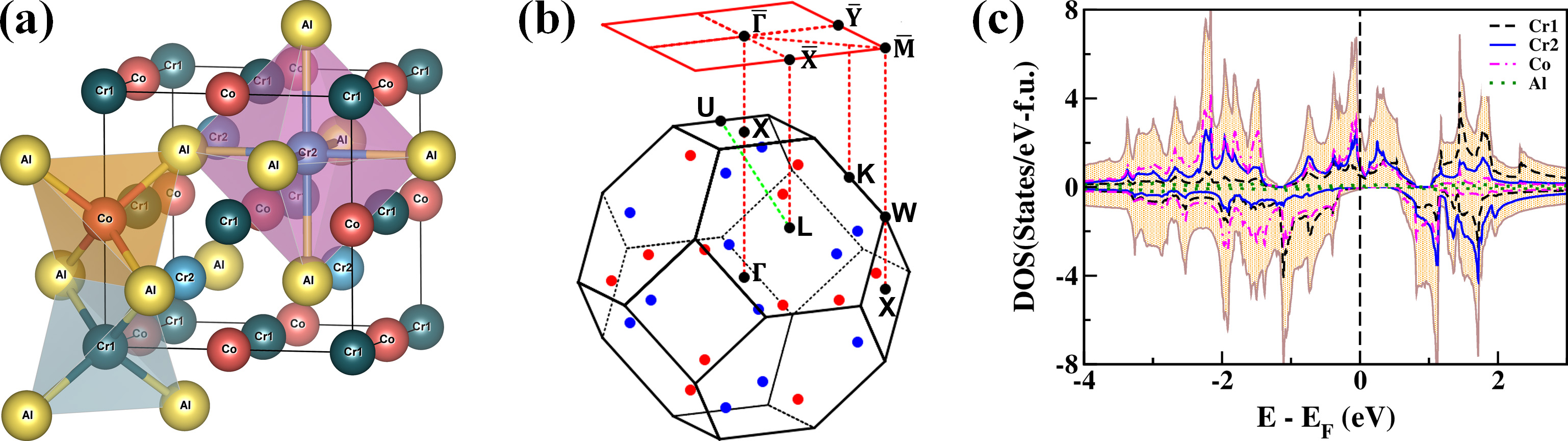}}}
\caption{
{\bf (a)} Structure of the inverse-Heusler \cca~
 with the sequence of Cr1-Cr2-Co-Al along the diagonal direction.
{\bf (b)} Bulk and (001) surface Brillouin zones (BZs) with high symmetry points
of the inverse Heusler systems.
The dots in the bulk BZ indicate Weyl points (WPs) with positive (red)
and negative (blue) chiralities.
There are twelve pairs of WPs protected by the three twofold rotational axes
along the $\langle100\rangle$ directions,
and six mirrors (${\cal M}_{\pm xy}$, ${\cal M}_{\pm yz}$, ${\cal M}_{\pm zx}$).
{\bf (c)} Total and atom-resolved densities of states (DOSs) of \cca, 
 showing the half-metallic character. The spin down DOS, with a gap at the Fermi
level $E_F$, is plotted downward.
}
\label{bz}
\end{figure*}

\section{Introduction and Background}
Over the past decade various topological phases in insulating, semimetallic,
 and even metallic materials 
have been proposed and intensively investigated due to the variety of exotic properties 
that emerge,
some of which have been experimentally realized.\cite{Chui2016,Yan2017}
More recently three-dimensional (3D) topological features mixing with 
zero-dimensional (0D) band-crossings
(Dirac, Weyl, multi-Weyl, and triple-nodal points [TNPs]) have stimulated 
further interest,\cite{arm18}
since Dirac and Weyl fermions have been sought among elementary particles, 
and conversely TNP fermions have no counterpart within the standard model. Nexus fermions are
a yet more intricate excitation that have been proposed. 

Breaking either parity ${\cal P}$ or time-reversal ${\cal T}$ symmetries in 3D systems,
or both, allows Weyl points (WPs)\cite{iTI} with topological character to appear. 
Weyl semimetal (WS) phases  were initially suggested, then observed, in transition metal monophosphides 
lacking ${\cal P}$ symmetry,\cite{sy_xu15,x.dai15,hasan15,she15,h.ding15} viz. TaAs.
This class shows an unconventional fermiology, with nodal loop Fermi surfaces\cite{ku15} in the bulk resulting 
in surface Fermi arcs connecting WPs of opposite chirality,\cite{x.dai15,hasan15,she15,h.ding15}
leading to unconventional transport properties such as large magnetoresistance and chiral anomaly effects.
In Weyl semimetals, the impact of breaking of ${\cal P}$ symmetry
depends on the strength of spin-orbit coupling (SOC)\cite{b.yan15} which often is small.
Partially for this reason, the magnetic Weyl (semi-)metals, which break ${\cal T}$, 
have begun to attract more interest.
In such magnetic materials WPs can appear even in the absence of SOC, 
and often show a much larger separation due to large spin polarization.

In addition to the exotic properties of ${\cal P}$ breaking WSs,
large anomalous Hall effects are expected in ${\cal T}$-broken cases\cite{felser16,ifw18}
where low carrier densities exist.
Since the number of pairs of WPs in a ${\cal T}$-broken semimetal is odd,\cite{iTI,z.wang16,y.jin17}
fewer WPs are likely than with the ${\cal P}$-broken cases.\cite{felser16,ifw18}
For example, a large anomalous Hall effect and angle, 
and a strong anomalous Nernst effect 
are proposed in the inverse-Heusler compensated half-semimetal Ti$_2$MnAl,\cite{ifw18,noky18} 
which has magnetic WPs just below the Fermi level and the same structure as \cca. 
Such a large anomalous Hall effect is also observed recently in a ferromagnetic 
van der Waals nodal line semimetal Fe$_3$GeTe$_2$.\cite{min18}
So far, only a few candidates have been predicted in the Co-based 
full Heuslers\cite{z.wang16,felser16},
half-metallic CrO$_2$,\cite{cro2_18} and the tetragonal 
$\beta$-V$_2$OPO$_4$,\cite{y.jin17} all of which do possess ${\cal P}$ symmetry. 
No magnetic Weyl phase has yet been observed, 
whereas a ${\cal P}$-broken WS was observed 
just a few months after the predictions.\cite{sy_xu15,h.ding15}

Another anomalous 0D band-crossing degeneracy is the TNP.
Initially, TNPs were predicted along the symmetry line with $C_{3v}$ symmetry
in both symmorphic and nonsymmorphic space groups,\cite{tnp1,tnp2}
and extended to lines of $C_{6v}$ and $C_{4v}$ symmetries.
In the $C_{4v}$ case, 
SOC removes the  6-fold degeneracy (considering spin)
and results in reverting to 4-fold Dirac or 2-fold Weyl nodes 
depending on existence of the ${\cal T}$ symmetry.\cite{vand18,ku_pbpd18}

An exotic phase has been expected when a 0D TNP coincides with a 1D nodal line.
This interconnection was dubbed nexus fermions by Chang {\it et al.}
who proposed it along the $C_{3v}$ symmetry line in tungsten carbide WC.\cite{hasan17}
However, that nexus point lies far above the Fermi level $E_F$ in WC.
Dispersion around a nexus point has similarities to the low energy excitations of
the chiral superfluid $^3$He-A,
as noted by Heikkil\"{a} and Volovik.\cite{volovik15}  
Chang {\it et al.} derived an unusual Landau level spectrum
quite distinct from that of Weyl semimetals,
suggesting novel magnetotransport response.\cite{hasan17} 
No realistic system has been proposed for such an exotic phase.

In this Rapid Communication we describe a unique material, Cr$_2$CoAl, that displays
{\it all four of the non-trivial degeneracies} mentioned above, as well as 
additional rare properties. Cr$_2$CoAl, with the non-centrosymmetric
inverse Heusler structure, is a ferrimagnetic metal displaying simultaneous
WPs, TNPs, nodal loops, and nexus fermions in the absence of SOC 
(which is minor due partially to cancellations).
It is furthermore a half metal, so the various fermionic degeneracies
mentioned above are pure spin. Finally, Cr$_2$CoAl is a rare 
compensated half-metal, producing no macroscopic magnetic field.
In the metallic spin-up channel the set of topologically features lie
close to $E_F$.

Due to a combination of space group symmetries in this system
the TNPs appear unexpectedly along the $C_{2v}$ symmetry line.
The Cr$_2$Co${\cal Z}$ systems (${\cal Z}$=Al, Ga, In) show Curie 
temperatures T$_C$ near 750 K, far above room temperature, and a
minute ordered moment (at most a few hundredths $\mu_B$),\cite{cca2}
confirming both strong magnetic coupling and magnetic compensation.
The Cr1 local moment is compensated by antialigned Co and Cr2 local 
moments;\cite{cca1,ku_cca18}
the Cr1 and Co sites form edge-sharing tetrahedra and 
Cr2 sites form octahedra, 
as displayed in Fig. \ref{bz}(a).
We will focus on \cca~ where correlation corrections reveal a half-metallic 
electronic structure.\cite{ku_cca18}
Our results suggest that several of these features may be common in 
magnetic inverse-Heusler systems.

\section{Calculational Methods}
Our {\it ab initio} calculations were based on 
the generalized gradient approximation\cite{gga} (GGA) implemented in
the all-electron full-potential code {\sc wien2k}.\cite{wien2k}
Some topological aspects were confirmed 
by another full-potential local-orbital code {\sc fplo}-18.\cite{fplo}
The same detailed conditions were used in these calculations as in our previous study.\cite{ku_cca18}
For these intermetallic compounds, correlation beyond GGA was excluded as its primary effect is
solely to widen the down-spin gap,\cite{correlation} producing a compensated half-metal
state without affecting the spin-up bands with the topological character.
These inverse-Heusler systems have 
the symmorphic but noncentrosymmetric $F\bar{4}3m$ (No. 216) space group, comprised
of all cubic point group operations except inversion.
Its bulk Brillouin zone (BZ) of $fcc$ shape is displayed in Fig. \ref{bz}(b).

The topological characters are investigated by the hybrid Wannier function 
charge center approach.\cite{hwcc}
From the band structure obtained from {\sc wien2k},
a tight-binding representation was generated in terms of maximally localized Wannier functions 
as implemented in the {\sc wannier90}\cite{w90} and {\sc wien2wannier}\cite{w2w} programs,
with an initial guess for the orbitals as the $3d$ orbitals of Cr and Co, and $3s$ and $3p$ orbitals of Al. 
The surface spectral functions were calculated by the Green's function approach,\cite{gf} 
implemented in the {\sc wanniertools} package.\cite{wt}


\begin{figure*}[tbp]
{\resizebox{16cm}{6cm}{\includegraphics{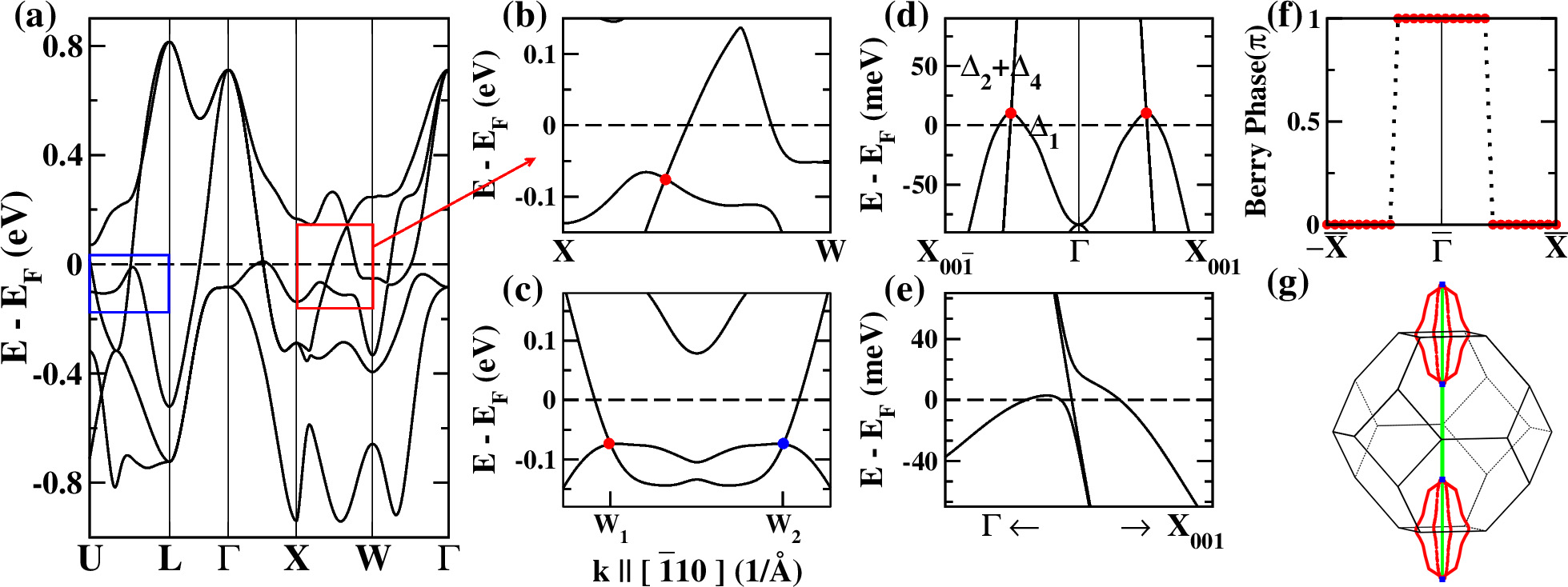}}}
\caption{
{\bf (a)} GGA spin-up band structure of \cca~ near $E_F$=0
in the absence of SOC. The spin-down bands (not shown) are gapped at $E_F$.
 The blue square indicates nodal points.
{\bf (b,c)} Enlarged spin-up plots around the Weyl points lying around --0.07 eV
along the $X-W$ line, and along a line parallel to the [$\bar{1}10$] 
direction, respectively.
   W$_1$ (W$_2$) denotes a WP with positive (negative) chirality
   at (0.4132,0,0.887)$\frac{2\pi}{a}$
     [(0,0.4132,0.887)$\frac{2\pi}{a}$].
{\bf (d)} Bands enlarged around the TNPs, lying just 10 meV above $E_F$ 
  along the [001] direction (SOC neglected).
  The red dots denote a pair of symmetry related TNPs, with the $\Delta_1$
  band having essentially zero velocity at the crossing.
 {\bf (e)} Effect of SOC on the TNP, showing a hybridization splitting of 10 meV, 
  for spin along the (001) direction.
 {\bf (f)} Plot of the $\pi$-Berry phase for the nodal loops, indicating 
  topologically nontrivial loops. 
  For the trivial nodal lines, the Berry phase is zero.
{\bf (g)} Plot of the nodal lines: a trivial along the $\Gamma-X$ rotational axis, 
  meeting with topologically nontrivial nodal loops centered on $X$. Only the 
  loops oriented along the $\hat z$ direction are shown. 
  Nexus fermions appear midway along $\Gamma-X$, marked by blue dots.
  TNPs (blue dots) occur at crossing points of two nodal lines lying on 
  two perpendicular mirror planes.}
\label{band}
\end{figure*}

\section{Topological character}
First we address topological properties of the compensated 
ferrimagnetic half-metal \cca, in the absence of SOC.
The weak effects of SOC will be discussed below.
Compared with the metallic spin-up bands,\cite{ku_cca18}
features of interest in the down bands (not shown) lie relatively far away 
from $E_F$ (lower than about --0.4 eV). 
Thus only the spin-up bands will be discussed in more detail.

Figure \ref{band}(a) displays the band structure near $E_F$.
In this region, the spin-up bands contain triplet $t_{1u}$ and doublet $e_{1u}$ manifolds 
driven by the hybridization among the $3d$ orbitals of Cr1, Co, and Cr2 ions, 
as previously discussed by some of the current authors.\cite{ku_cca18}
Along each of the twelve $X-W$ lines in the BZ, 
a linear crossing occurs below E$_F$ at $E_{WP}=-74$ meV.
To reveal the origin of the nodal points, their chiralities are calculated 
by the Wannier charge center approach implemented in the {\sc wanniertools} package.\cite{wt}
Figure \ref{band}(c) shows two WPs with opposite chirality along a line parallel to the [$\bar{1}10$] direction.
In total, there are twelve pairs of WPs protected by the six mirror planes
and three two-fold rotation symmetries, as displayed on the bulk BZ in Fig. \ref{bz}(b).
The positions and chiralities are given in the caption of Fig.~\ref{band}. 
As mentioned above, 
magnetic Weyl semimetals with ${\cal P}$ symmetry have an odd number 
of pairs of WPs,\cite{iTI}
but this system has an even number due to lack of both 
${\cal T}$ and ${\cal P}$ symmetries.

A second feature is the band crossing 
between doublet and singlet bands just above E$_F$ at E$_{TNP}$=10 meV
along the $\Gamma-X$ line having $C_{2v}$ symmetry, enlarged in Fig. \ref{band}(d).
These Weyl touchings do not fit well with the types of Ghang {\it et al.}\cite{hasan17} 
who classified crossings as of the same sign, or opposite signs. This band touching
accidentally occurs when one band has zero velocity: static fermions crossing and mixing
with Weyl fermions.

In this cubic system, there are three pairs of TNPs. (For the origin of the TNPs, see below.)
Each TNP lies midway along the $\Gamma-X$ line.
The doublet bands along (0,0,$k_z$) have mostly $d_{xz}$, $d_{yz}$ 
character of the Co and Cr1 ions, 
whereas the singlet band has mainly $d_{z^2}$ character of the Co and Cr2 ions,
with some mixing of Cr2 $d_{xy}$.

This crossing, however, leads to a third unexpected feature. 
Around --0.1 eV midway along the $U-L$ line two bands cross, one again
with essentially zero velocity.
Unusually, in one direction perpendicular to this line, the band touching persists.
Nodal line calculations, using the {\sc wanniertools} code,\cite{wt} 
establish that the crossing leads to two intersecting nodal lines on the 
two perpendicular mirror planes 
about the $C_2$ rotational axis, as given in Fig. \ref{band}(g). 
Analysis of the Berry phase\cite{allen07,allen18,hasan16}
or the behavior of hybrid Wannier charge centers\cite{h.huang18} can clarify
the occurrence and topological nature of a nodal line.

One way is by calculating the  Berry phase by integrating around a closed loop 
in the BZ,\cite{hasan16,x.zhang17} as shown in Fig. \ref{band}(f).
The integral vanishes (modulo 2$\pi$)
unless it encircles
a nontrivial nodal line. This method even allows, 
for topological loops with Zeeman (magnetic) band splitting,
 the detection of the  nodal lines after splitting by
SOC.\cite{allen07,allen18} 
Alternatively, 
the Berry phase resulting from
the sum of the hybrid Wannier charge centers $z(\vec k_{\parallel})$  
shows a jump when $z(\vec k_{\parallel})$ crosses the 
projection of a topological
nodal loop.\cite{h.huang18} For this noncentrosymmetric space group the Berry
phase is no longer quantized, but the topology-revealing jump still occurs.
  (See the Supplementary Material.)\cite{supp}
The TNPs lie at the crossing points of these $X$-centered nodal loops and 
a trivial nodal line along the rotational axis imposed by crystal symmetry,
resulting in the curious nexus fermionic region.
With quadratically touching bands, the 1D nodal line along $\Gamma-X$ is 
topologically trivial.\cite{tnp2,hasan17}

\begin{figure*}[htbp]
{\resizebox{16cm}{9cm}{\includegraphics{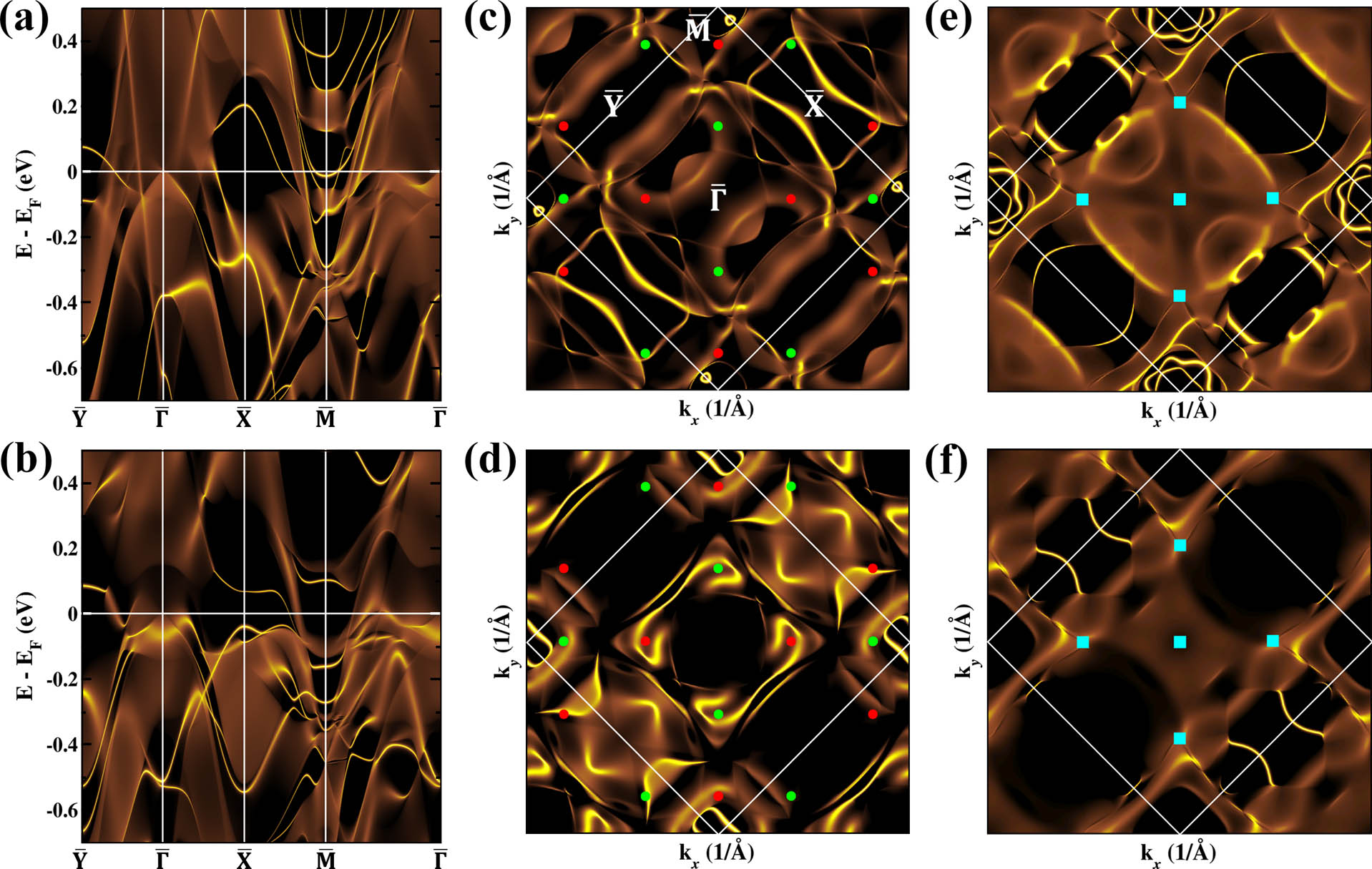}}}
\caption{
Several views of surface spectra neglecting SOC, with bright 
yellow indicating
high surface intensity. 
Top and bottom rows of panels are for the Cr1-Co and Cr2-Al 
surface terminations, respectively.
(a)--(b): The (001) surface spectral functions of the spin-up channel 
along symmetry lines.
These are followed by isoenergy spectral densities of the surface states 
at energy lying at the WPs (middle column) and TNPs (right column). 
The WPs in (c) and (d) are denoted as green and red circles, 
whereas the light-blue squares of (e) and (f) indicate TNPs.
The $\bar{X}$ and  $\bar{Y}$ points in (a) and (b) lie at each midpoint of adjacent faces
of the surface BZ,
outlined by white lines in panels (c)--(f).
The high-symmetry points relative to bulk are provided in Fig. \ref{bz}(b).
}
\label{surf}
\end{figure*}

\section{Origin of triple nodal points}
TNPs occur most commonly along symmetry lines when a nondegenerate
band crosses a doubly degenerate band. Since the band eigenfunctions belong to different
irreducible representations of the little group, there is no matrix element of the
Hamiltonian (which has the full symmetry of the crystal) to mix the bands, and they
cross. Nevertheless, the crossing causes the eigenset at the TNP to become
non-analytic, leading to possible topological character.
(See the Supplementary Materials.)\cite{supp}

In this magnetic and noncentrosymmetric inverse-Heusler material,
TNPs can arise along the $\Gamma-X$ lines 
(in the absence of SOC).
Such TNPs along the $\Gamma-X$ line of the $C_{2v}$ point group 
have not been considered before. 
The TNPs 
along the $\Gamma-X$ line result from the conventional little group
symmetries along the $\Delta$ line in the $F{\bar 4}3m$ space group.

\section{Surface states}
Figures \ref{surf}(a) and \ref{surf}(b) show the (001) surface spectral 
functions along symmetry lines
for the two terminations Cr1-Co and Cr2-Al, respectively.  
The mapping from the bulk BZ onto the surface BZ is presented in Fig. \ref{bz}(b).
Note that the spectra along the
$\bar{\Gamma}-\bar{X}$ and $\bar{\Gamma}-\bar{Y}$ lines 
are asymmetric due to surface termination breaking of square symmetry.
This asymmetry is also reflected in the isoenergy spectral densities shown 
throughout the surface BZ Figs. \ref{surf}(c)--\ref{surf}(f),
where their structures are identical only along the diagonal directions.

Near $E_F$, several surface states are visible for both terminations.
Topological nodal lines lead to drumhead surface states 
within the projection of the loop,
with (usually) low dispersive surface states,\cite{burkov11,hasan16}
suspected to support instabilities toward surface superconductivity or magnetism
when they lie near $E_F$.
(Drumhead states in isoenergy spectra appear as closed contours within
projection of the nodal loop, or as lines terminating at the edge of the projection.)
Along the $\bar{\Gamma}-\bar{Y}$ line, 
the weak drumhead related states lying near $E_F$ appear around 
--75 meV (+75 meV) for the Cr1-Co (Cr2-Al) termination, 
as shown in Figs. \ref{surf}(a) and \ref{surf}(b).
However, the main nodal loop projection is along the $\bar{\Gamma}-\bar{M}$ (see below).

Figures \ref{surf}(c) and \ref{surf}(d) show the isoenergy spectral densities of the two terminations
at the WP energy $E_{WP}=$--74 meV.
Fermi arcs can be seen connecting WPs with opposite chirality.
We also show the spectral densities at the TNP energy $E_{TNP}=10$ meV
in Figs. \ref{surf}(e) and \ref{surf}(f).
In addition to the one projecting onto $\bar{\Gamma}$, TNPs are projected near the 
midpoint of $\bar{\Gamma}-\bar{M}$ lines.
In the Cr1-Co termination, Fermi arcs connecting pairs of TNPs are visible,
losing intensity as they merge into the bulk spectrum.
At the WP energy, Fig. \ref{surf}(c), arcs extend between TNPs without merging 
into bulk bands. 
The nearness of the TNPs to $E_F$ make them amenable to measurement.
We are not aware of any topological invariant involving TNPs having been identified
in a real material.\cite{arm18}

\subsection{Effects of spin-orbit coupling}
Lowering of symmetry and lifting of degeneracies by SOC depend 
on the direction of magnetization,
thus the (still near) topological character can be anisotropic.
Effects of SOC on WPs in inverse-Heusler compounds
have been analyzed by Shi {\it et al.},\cite{ifw18} where they establish
the minor effect for most purposes.
We calculate the magnetic anisotropy energy of Cr$_2$CoAl to be at most
4 meV/f.u., with the (111) direction favored slightly. 
Just as the spin moments exactly cancel due to antialigned atomic moments,
the SOC-driven orbital moments and change in spin moments, already small,
also cancel, with the result $\mu_{s}=0.01, \mu_{orb}=-0.02,
\mu_{net}=-0.01,$ in $\mu_B$.

Figure \ref{band}(e) shows a closeup view of the GGA+SOC band structure 
near the TNPs along the $\Gamma-X_{001}$ line with spin along (001). 
SOC leads to a hybridization splitting of about 10 meV at the TNPs, but
affects the Fermi surface very little. 
This mixing strength is minor compared with the observed high Curie temperature 
of $k_B$T$_C\approx$ 65 meV.  Kim {\it et al.} have shown
that a tiny gap, comparable with the size of thermal fluctuation $E_{TNP}/k_B$, 
results in surface spectra and transport properties similar to those without SOC.\cite{vand18}



\section{Summary}
We have also studied the isovalent and isostructural Ga and In analogs.
In addition to WPs,
these systems also show nexus fermions very near $E_F$ midway 
along the $\Gamma-X$ line in the spin-up channel. 
The energies are --6 (--190) meV for Cr$_2$CoGa (Cr$_2$CoIn).
(See the Supplementary Material.)\cite{supp}

In summary, using first principles calculations
we have uncovered a unique combination of topological character and compensated 
half-metallic magnetic order in the noncentrosymmetric, time-reversal symmetry breaking
inverse Heusler compound \cca.
Directly associated with the lack of ${\cal P}$ and ${\cal T}$ symmetries,
\cca~ displays a combination of four unusual degeneracies: magnetic 
Weyl points, triple nodal points, both topological and
trivial nodal loops that interconnect, and nexus fermions. 
All of these occur in a half metal with compensating magnetic moments,
which provide no macroscopic magnetic field that would complicate some probes. 
Specifically, the gapped spin-down electrons will not interfere with the spin-up
topological features within the gap. 
The weak SOC in $3d$ metals leads
to tiny orbital moments and band shifts that are negligible for most purposes.

Unprecedented TNPs emerge along the $C_{2v}$ ({\it i.e.}, $\Gamma-X)$ line 
due to a combination of the rotation and mirror point group symmetries.
These TNPs along the $\Gamma-X$ line interconnect with nodal links on the mirror planes, 
leading to nexus fermions lying right above the Fermi energy in the spin-up channel.
The combination of compensated half-metallicity and nexus points very near 
the Fermi energy,
with high Curie temperature and minor SOC effects, 
makes \cca~a promising candidate to realize an observable nexus fermion phase
using modern spectroscopies and transport studies.
Many of these features should be generic in inverse-Heusler magnets, 
with energetic positions depending on the specific compound.

\section{Acknowledgments}
We acknowledge M. Richter, K. Koepernik, U. R\"o$\ss$ler, 
R. Ray, and J. Facio 
for fruitful discussions, and Y. Kim for a useful 
discussion on triple nodal points.
This research was supported by NRF of 
Korea Grant No. NRF-2016R1A2B4009579 (H.S.J, Y.J.S, and
K.W.L), and by NSF DMREF Grant DMR-1534719 (W.E.P).
K.W.L gratefully acknowledges the hospitality of 
IFW Dresden during his sabbatical.

\newpage
\section{Supplemental Material}

\section{Characters of triple nodal points and lines}
\begin{figure*}[htbp]
{\resizebox{12cm}{7.5cm}{\includegraphics{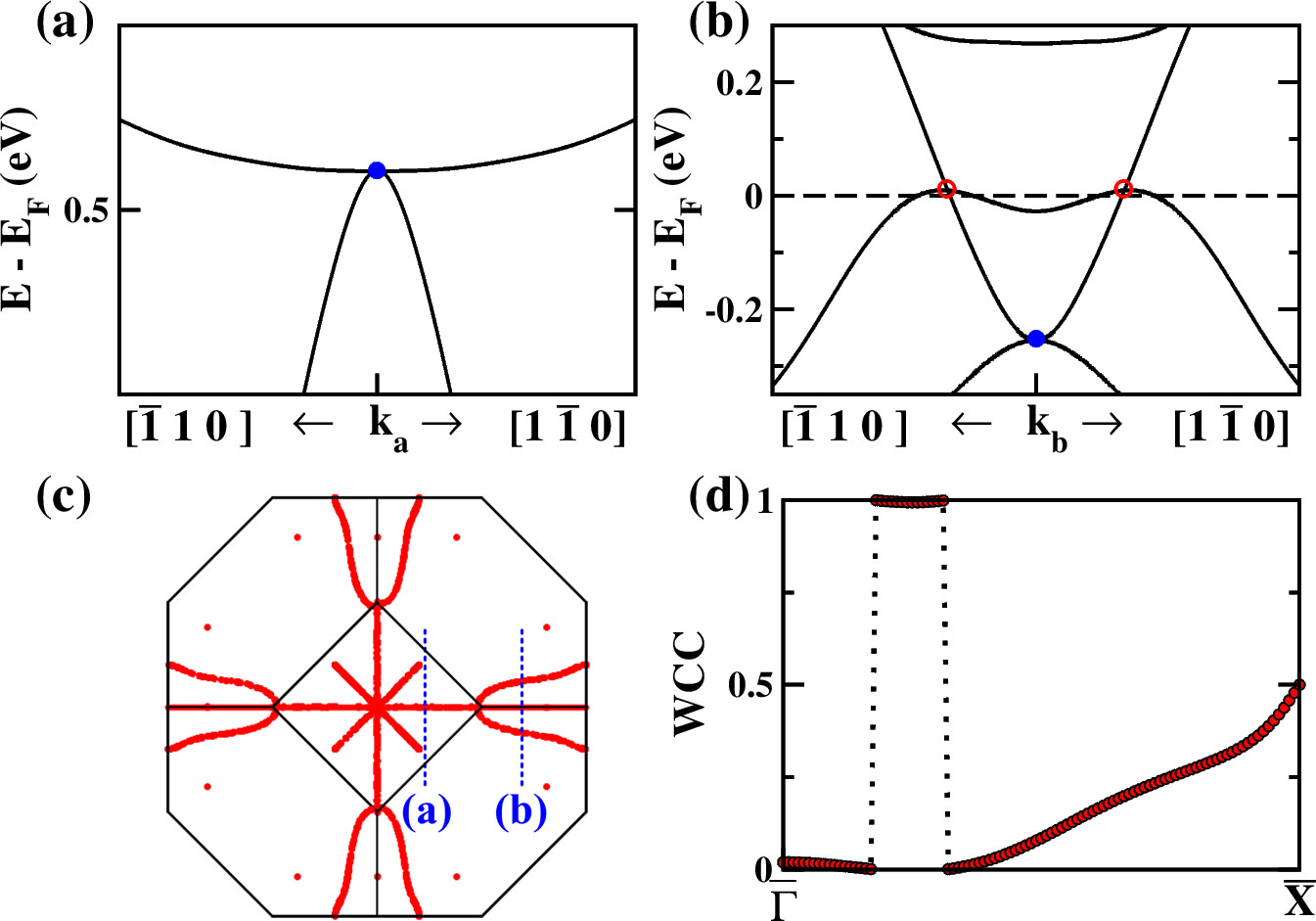}}}
\caption{(a),(b) Band crossings around nodal points at $\vec{k}_a=(0,0.1153,0.1153)$ and
$\vec{k}_b=(0,0.3458,0.3458)$ in units of $\frac{2\pi}{a}$. 
The (blue) filled circles denote quadratic crossings of the type-$\alpha$ (trivial) nodal lines,
whereas the (red) circles mark linear crossing of the type-$\beta$ 
(nontrivial) nodal loops on the mirror planes.
The paths are described by dashed lines in (c),
which is (001) projection of the bulk Brillouin zone.
The (red) solid lines indicate some of nodal lines and loops.
(d) Plot of the hybrid Wannier charge center WCC $z(\vec k_{\parallel})$ of the nodal loops of (c)
  for $\vec k_{\parallel}$ along the indicated path, showing the jump in Berry phase as 
  the WCC passes through the
  projection of the topological nodal loop. 
}
\label{s1}
\end{figure*}

Figure~\ref{s1} provides more detail about the nodal points and lines discussed in
the main text; see the caption for explanation. 

\begin{figure*}[htbp]
{\resizebox{12cm}{7.5cm}{\includegraphics{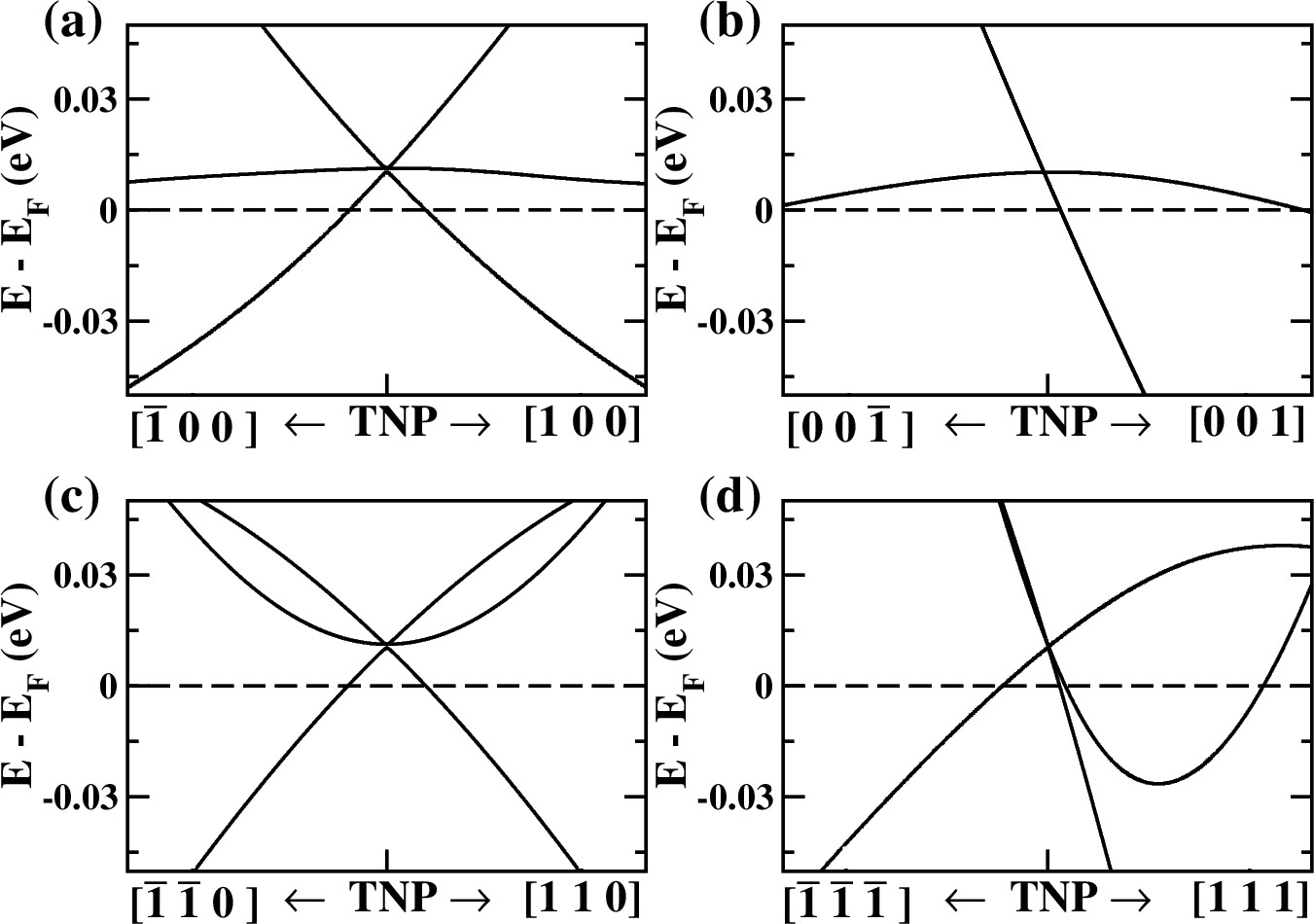}}}
\caption{Band dispersions around the non-analytic TNP in Cr$_2$CoAl,
illustrating the various types of dispersion that eminate from the TNP at (0.2527,0.2527,0)$\frac{2\pi}{a}$.
}
\label{s2}
\end{figure*}

The unusual variations in band dispersion at emerge from the topological TNP
are shown in Fig.~\ref{s2}.

\section{Band structures and surface spectra of the G\lowercase{a} and I\lowercase{n} analogs}

\begin{figure*}[tbp]
\vskip 10mm
{\resizebox{12cm}{6cm}{\includegraphics{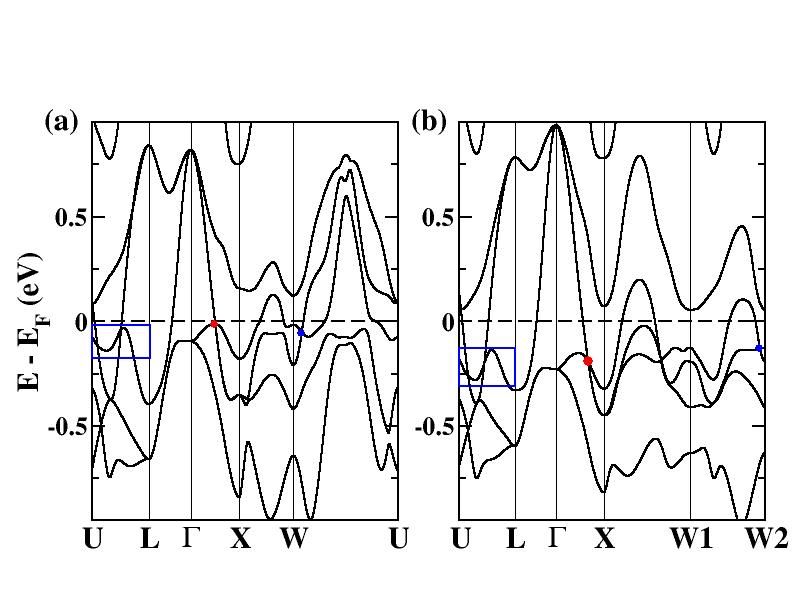}}}
\caption{GGA blowup spin-up band structures near $E_F$ of (a) Cr$_2$CoGa and (b) Cr$_2$CoIn,
  which are similar to that of \cca, in the absence of SOC.
 The red (blue) dots denote triple (Weyl) nodal points,
 while the (blue) box indicates nodal points leading to nodal line on each mirror planes.
 In (b), the positions of W1 and W2 points are given by
 ($\frac{1}{2}$,$\frac{1}{4}$,$-\frac{1}{4}$) and ($\frac{1}{4}$,$\frac{3}{4}$,$\frac{1}{2}$), 
 respectively, in units of $\frac{2\pi}{a}$.
}
\label{s3}
\end{figure*}

Figure~\ref{s3} provides the bands near E$_F$ for the isostructural and isovalent
compounds Cr$_2$CoGa and Cr$_2$CoIn, showing how the different energy positions
(``different chemistry") affect the various band crossings.

\begin{figure*}[htbp]
{\resizebox{16cm}{9cm}{\includegraphics{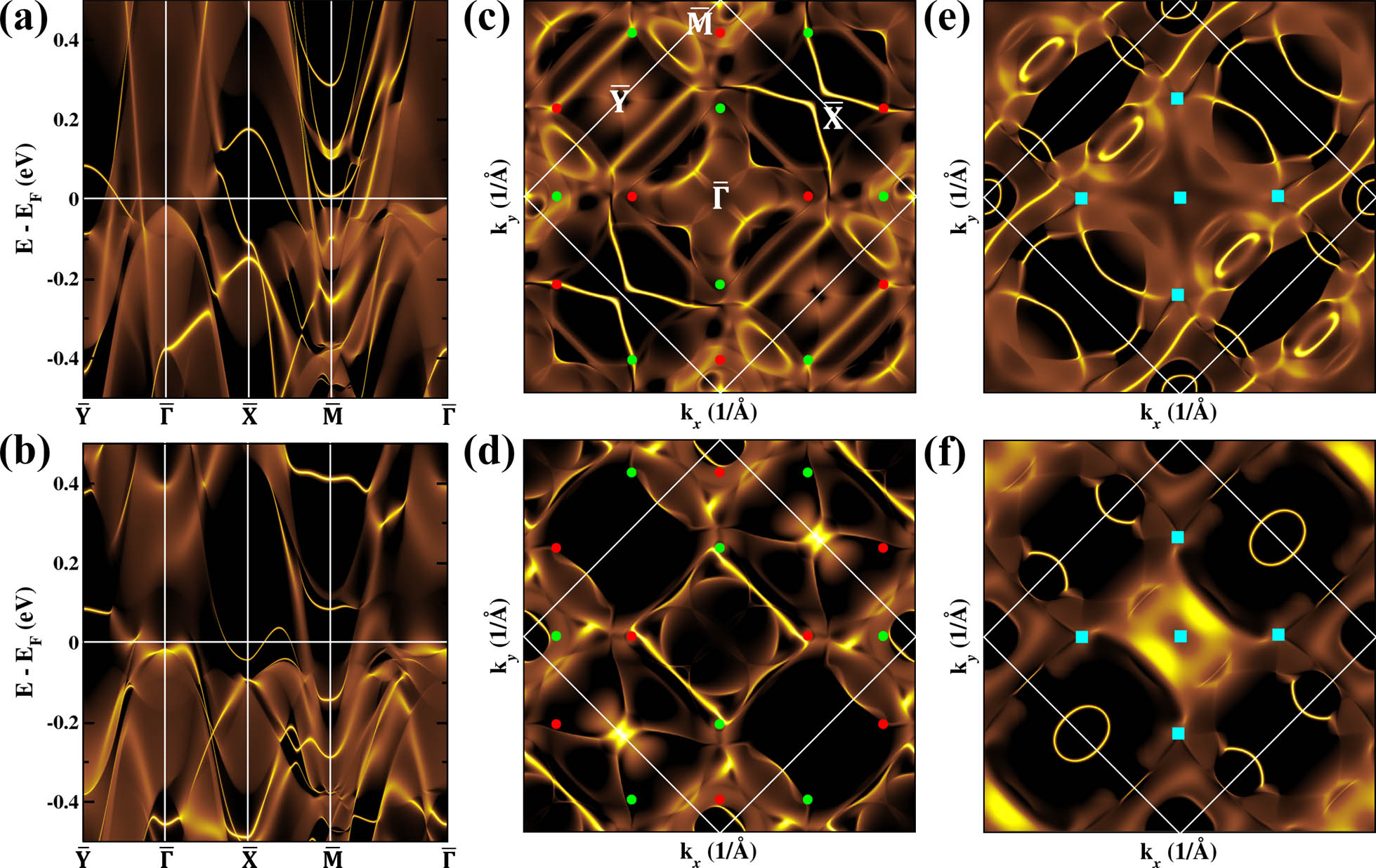}}}
\caption{Surface spectra of Cr$_2$CoGa, neglecting SOC, 
with bright yellow indicating high surface intensity. 
Top and bottom rows of panels are for the Cr1-Co and Cr2-Ga
surface terminations, respectively.
(a), (b): the (001) surface spectral functions of the spin-up channel 
along symmetry lines.
These are followed by isoenergy spectral densities of the surface states 
lying at the WP energy $E_{WP}$=--48 meV (middle column) and at the TNP
energy $E_{TNP}$=--6 meV (right column). 
The WPs in (c) and (d) are denoted green and red circles, 
whereas the light-blue squares of (e) and (f) indicate TNPs.
The $\bar{X}$ and  $\bar{Y}$ points in (a),(b) lie at each midpoint of adjacent faces
of the surface BZ, outlined by white lines in panels (c)--(f).
}
\label{s4}
\end{figure*}

\begin{figure*}[htbp]
{\resizebox{16cm}{9cm}{\includegraphics{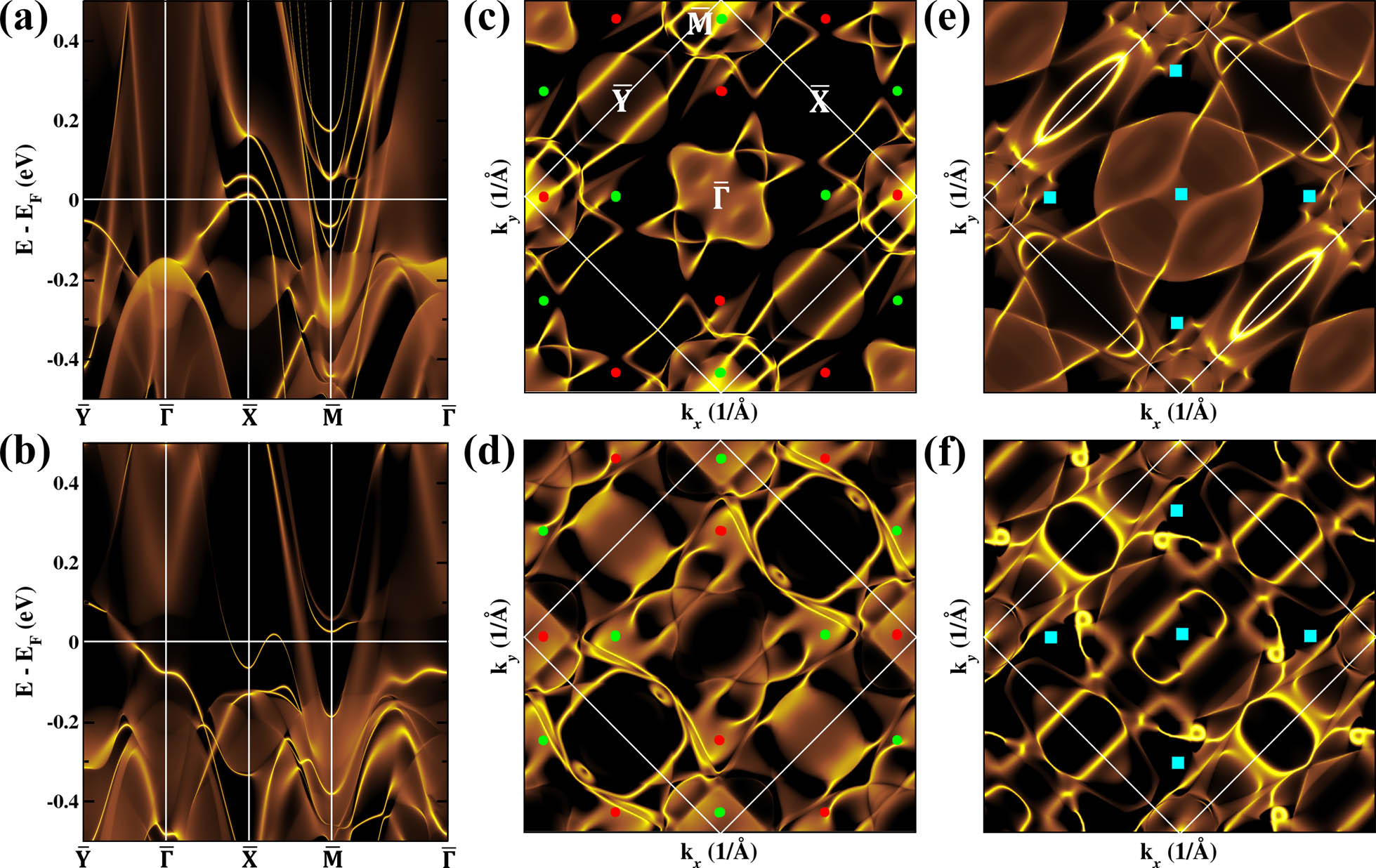}}}
\caption{Correspondence of Fig. \ref{s4} for Cr$_2$CoIn.
Top and bottom rows of panels are for the Cr1-Co and Cr2-In 
surface terminations, respectively. The corresponding topological points
lie at $E_{WP}$=--130 meV, $E_{TNP}=$--190 meV.
}
\label{s5}
\end{figure*}

Figures \ref{s4} and \ref{s5} show the surface spectral functions for Cr$_2$CoGa and Cr$_2$CoIn,
respectively, presented as for Cr$_2$CoAl in Fig. 3 of the main text. See the captions
for descriptions of the various panels.


\begin{thebibliography}{10}
\bibitem{correlation} 
As shown in Ref. \cite{ku_cca18}, correlation leads to a slight shift of the band structure.
Thus, variation of correlation strength shift only positions of the topologically nontrivial points discussed here.
\end{thebibliography}
\end{document}